\documentclass[fleqn,12pt]{article}
\usepackage{espcrc1}
\usepackage {amsfonts}
\usepackage {amsmath}
\usepackage {latexsym}
\usepackage{graphicx}
\newcommand{\bg}[1]{\mbox{\boldmath $#1$}}

\title{MICROSCOPIC DYNAMICS OF HARD ELLIPSOIDS IN THEIR LIQUID AND
  GLASSY PHASE} 

\author{A. Latz $^1$, M. Letz $^2$, R. Schilling and Th.
Theenhaus  
\address{Institut f\"ur Physik, Johannes Gutenberg Universit\"at, Mainz
D-55099 Mainz, Staudinger Weg 7, Germany\\
$^1$ Institut f\"ur Physik, Reichenhainer Strasse 70,
TU-Chemnitz, D-09107 Chemnitz, Germany\\
$^2$ Schott Glas, Research and Development, Hattenbergstr. 10,
D-55014 Mainz, Germany}}

\begin{document}

\maketitle

\begin{abstract}
To investigate the influence of orientational degrees of freedom
onto the dynamics of molecular systems in its supercooled and
glassy regime we have solved numerically the mode-coupling
equations for hard ellipsoids of revolution. For a wide range of
volume fractions $\phi$ and aspect ratios $x_{0}$ we find an
orientational peak in the center of mass spectra
$\chi_{000}^{''}(q,\omega)$ and $\phi_{000}^{''} (q,\omega)$ about one
decade below a high frequency peak. This orientational peak is the
counterpart of a peak appearing in the quadrupolar spectra
$\chi_{22m}^{''}(q,\omega)$ and $\phi_{22m}^{''}(q,\omega)$. 
The latter peak is almost insensitive on
$\phi$ for $x_{0}$ close to one, i.e. for weak steric hindrance,
and broadens strongly with increasing $x_{0}$. Deep in the glass
we find an additional peak between the orientational and the high
frequency peak. We have evidence that this intermediate peak is
the result of a coupling between modes with $l=0$ and $l=2$, due
to the nondiagonality of the static correlators.

\end{abstract}

\section{INTRODUCTION}

Although experimental data on plastic crystals \cite{1,2,3} have
demonstrated that orientational degrees of freedom by themselves
show glasslike dynamics and even can undergo a glass transition
their role for the dynamics and the glass formation of supercooled
liquids has not been studied in great detail. With the extension
of mode-coupling theory (MCT) from simple liquids \cite{4,5,6,7}
to molecular systems \cite{8,9,10,11} there is a
\textit{microscopic} theory which allows to investigate the role
of the rotational degrees of freedom and their coupling to the
translational ones. It is desirable to choose systems for which
this coupling can be varied systematically.  From this point of
view hard ellipsoids of revolution represent an interesting model
system. Keeping their minor axis $b$ fixed, a change of the aspect
ratio $x_{0}=a/b$, where $a$ is the major axis, will change the
steric hindrance between the ellipsoids. This again will result in
a change of the translational-rotational coupling. Applying MCT to
hard ellipsoids it was shown that for $x_{0}$ about two or larger
the ellipsoids freeze into a glass at a critical volume fraction
$\phi_{c}(x_{0})$ due to the appearance of a medium range 
orientational order which originates from a precursor
of a nematic order \cite{12}. Recently we have shown that this
precursor of nematic order also has an influence onto the
dynamical features in the microscopic time and frequency regime
\cite{13}. From both, the solution of the corresponding
MCT-equations and from a MD-simulation a peak in the
compressibility spectrum has been found at $\omega_{op}$ about one
decade below a high-frequency peak (hf-peak). The additional peak
clearly has an orientational origin since $\omega_{op}$ scales
with $I^{-1/2}$, $I$ being the moment of inertia with respect to
the minor axis. Therefore we called it an orientational peak $(op)$.
These calculations were restricted to a single aspect ratio,
$x_{0}=1.8$, and one volume fraction, only. It is the motivation
of the present contribution to study the microscopic dynamics and 
in particular the $op$ as a function of $x_{0}$ and $\phi=\frac{\pi}{6}
\varrho_{0}x_{0}$ where $\varrho_{0}$ is the number density. 
In this respect we would also like to mention that studying
the \textit{athermal} system of hard ellipsoids can also be
instructive for real molecular liquids composed of linear
molecules like e. g. diatomic molecules. Indeed, it has been
demonstrated that the static correlators $S_{lm,l'm'}(\vec{q})$ and
the non-ergodicity parameters $f_{lm,l'm'} (\vec{q})$ for diatomic
molecules with Lennard-Jones interactions obtained from a
MD-simulation \cite{14} can be well reproduced by a system of hard
ellipsoids with $x_{0}=1.5$ \cite{15}. Note that this aspect ratio was not
the result of fitting procedure, but could be determined from the
Lennard-Jones parameters.

Besides hard ellipsoids MCT has already been applied to study the
dynamics of a dumbbell-molecule in an isotropic liquid of hard
spheres \cite{16}, \cite{17} and of a liquid of dipolar hard spheres
\cite{18}. Like for hard ellipsoids, the change of the elongation
$\zeta$ of the dumbbell allows to vary the steric hindrance with
the hard sphere environment. This effect has been investigated
quantitatively for the orientational correlators $C_{l}
(t)=\langle P_{l}(\vec{e}(t)\cdot {\vec{e}(0))}\rangle $ where
$P_{l}$ are the Legendre Polynomials and $\vec {e}(t)$ is the unit
vector pointing along the symmetry axis of the dumbbell at time
$t$. It has been found that weak and strong steric hindrance can
lead to quite different relaxational behavior. For instance, in
case of sufficiently small elongations the $\alpha$-relaxation
scaling law is strongly disturbed. The correlator with odd-$l$ and
even-$\l$ behave different. It is interesting that this finding
agrees with that found from a MD-simulation of a diatomic
Lennard-Jones liquid. \cite{19}.

For dipolar hard spheres it is the dipolar interaction which is
responsible for a translational-rotational-coupling. This type of
coupling is different to that generated by steric hindrance.
Nevertheless, we have also found in the compressibility spectrum
an $op$ about a decade below a hf-peak \cite{18}. This may suggest that an
$op$ can be a rather general feature of a molecular liquid. This
motivates to study for hard ellipsoids the microscopic dynamics as
a function of the aspect ratio and volume fraction.

The organization of our paper is as follows. In the next section
we will introduce the basic quantities and their equations of
motion. The results from a numerical solution of these equations
will be presented and discussed in section III and in section IV
we will summarize and will give some conclusions.

\section{Equations of motion: Hard ellipsoids}

We consider a system of $N$ hard ellipsoids of revolution with
mass $M$ and moment of inertia $I=M(1+x_{0}^{2})/20$ with respect
to the minor axis. $x_{0}=a/b$ is the aspect ratio, i.e. the
ratio of the major and minor axis. The length and time unit is
chosen such that $b=1$ and $M=1$. The microscopic configurations
are given by the center of mass position $\vec{x}_{n}$ and the set
of Euler angles $\Omega_{n}=(\varphi_{n}, \vartheta_{n},
\chi_{n})$, $n=1,2,\ldots, N$. Because of the rotational
invariance around the symmetry axis of the ellipsoids of
revolution the third Euler angle $\chi_{n}$ is not involved in the
steric hindrance. Consequently this degree of freedom will not be
taken into account. Then, the basic quantities for the study of
the molecular dynamics are the \textit{tensorial} density
correlators

\begin{equation} 
\label {eq1}
S_{lm,l'm'}(\vec{q},t)= \frac{1}{N} \langle \rho_{lm}^{*}
(\vec{q},{t}) \rho_{l'm'}(\vec{q},{0}) \rangle
\end{equation}

where $\rho_{lm}(\vec{q},t)$ is the microscopic tensorial
density determined by 
$\{ \vec{x}_{n}(t) \}$ and $\{ \varphi_{n}(t),\,
\vartheta_{n}(t)\}$. $l=0,1,2,\ldots$ and $-l\leq m \leq l$ specify
the rotational part of the density modes. For completeness we
mention that the dynamics of molecular liquids can also be
described by a site-site formalism \cite{9}. However, hard
ellipsoids are not made up for a finite number of sites. Hence one
has to approximate a hard ellipsoid by, e. g. a two-site dumbbell
where each site is the center of a sphere as it was used in refs.
\cite{16}, \cite{17}.

The correlation matrix $\mathbf{S}(\vec{q},{t})=(S_{lm,l'm'}
(\vec{q},{t}))$ simplifies as follows. First of all we can choose
the $q$-frame, i.e. it is $\vec{q}=(0,0,q)$ , $q=|\vec{q}|$. This
makes $\mathbf{S}(\vec{q},t)$ diagonal in $m$ \cite{10}:

\begin{equation} \label {eq2}
S_{lm,l'm'}(\vec{q},t) = S_{ll'm}(q,t) \delta_{mm'}\;.
\end{equation}

In addition one can prove \cite{10} that $S_{ll'm}(q,t)$ are real
and that

\begin{equation}
\label{eq3}
S_{ll'-m}(q,t)=S_{ll'm}(q,t)\,.
\end{equation}

Second the head-tail symmetry of the ellipsoids implies that the
collective correlators $S_{lm,l'm'}(\vec{q},t)$ vanisch for $l$
and/or $l'$ odd. For its self part $S_{lm,l'm'}^{(s)}(\vec{q},t)$,
which will not be considered here, this is only true for $l+l'$
odd.

How to derive equations of motion for $\mathbf{S}(\vec{q},t)$ has been
described in great detail \cite{10}. Here we repeat just the
result:

\begin{subequations}
\label{eq4}
\begin{eqnarray} 
\label{eq4a}
 & \dot{\mathbf{S}} (\vec{q},t) + i \sum\limits_{\alpha=T,R}
\mathbf{q}^{\alpha} \mathbf{S}^{\alpha} (\vec{q},t)  =0 & \\[4ex]
\label{eq4b}
& \dot{\mathbf{S}}^{\alpha}(\vec{q},t) + i\mathbf{q}^{\alpha}
\mathbf{J}^{\alpha}(\vec{q}) \mathbf{S}^{-1}
(\vec{q})\mathbf{S}(\vec{q},t) & 
\nonumber \\
& +\mathbf{J}^{\alpha}(\vec{q})
\int\limits_{0}^{t} dt'
\sum\limits_{\alpha'=T,R}\mathbf{m}^{\alpha \alpha'}
(\vec{q},t-t')\mathbf{S}^{\alpha'}(\vec{q},t')  =0 &
\end{eqnarray}
\end{subequations}

with the initial conditions:

\begin{align}
\mathbf{S}(\vec{q},0) & \equiv \mathbf{S}(\vec{q})\, , &  
\mathbf{S}^{\alpha}(\vec{q},0) & \equiv 0 \, ,
\nonumber \\
\label{eq5}
\dot{\mathbf{S}}(\vec{q},0) & \equiv 0 \, , & 
\dot{\mathbf{S}}^\alpha (\vec{q},0) & 
\equiv -i\mathbf{J}^{\alpha}(\vec{q})\mathbf{q}^{\alpha}
\end{align}

and

\begin{equation} 
\label{eq6}
(\mathbf{q}^{\alpha})_{lm,l'm'}= \delta_{ll'} \delta_{mm'}
\left\{ 
\begin{array}{cc}
q \, & \alpha = T \\
\sqrt{l(l+1)} \, & \alpha =R
\end{array}
\right.
\end{equation}

\begin{equation} 
\label{eq7}
(\mathbf{J}^{\alpha} (\vec{q}))_{lm,l'm'} 
= (k_{B} T / I_{\alpha}) \delta_{ll'} \delta_{mm'}
\end{equation}

where $I_{T}=M$ and $I_{R}=I$. $\mathbf{S}(\vec{q})$ and
$\mathbf{J}^\alpha  (\vec{q})$ are the static density and current
density correlators, respectively. The form of this set of
equations is different but equivalent to that derived in ref.
\cite{10}. The main difference is the occurrence of the density -
current density correlators $\mathbf{S}^\alpha (\vec{q},t)$ in
Eqs.~(\ref{eq4}). It has been demonstrated that the set of equations
~(\ref{eq4}) is appropriate for a numerical solution \cite{18}.
Eqs.~(\ref{eq4}) become closed by the MCT-approximation for the
memory kernels $m_{lm,l'm'}^{\alpha\alpha '} (\vec{q},t)$. With the
shorthand notation $\lambda={lm}$ it is \cite{10}

\begin{eqnarray}
m^{\alpha \alpha '}_{\lambda \lambda '} (\vec {q}, t) =
\frac {1}{2N} \sum _ {\vec {q}_{1} \vec {q}_{2}
} \sum _{\lambda _{1} \lambda'_{1} \atop \lambda _{2} 
 \lambda'_{2}} \Big( \sum _ {\lambda ''} [u^{\alpha} (\vec
 {q} \lambda | \vec {q}_{1} \lambda '' ; \vec
 {q}_{2} \lambda _{2}) c_{\lambda ^{''}\lambda _{1} }(\vec
 {q}_{1}) + (1 \leftrightarrow 2)]) \Big) \cdot \nonumber \\
  \cdot \Big( \sum_{\lambda '''}[u^{\alpha'} 
 (\vec{q} \lambda^{'} | \vec {q}_{1} \lambda {'''} ; \vec
 {q}_{2} \lambda _{2}^{'}) c_{\lambda ^{'''}\lambda _{1}^{'} }(\vec
 {q}_{1}) + (1 \leftrightarrow 2)]\Big)^{*} 
 S_{\lambda_{1} \lambda'_{1}}
 (\vec{q}_{1},t) S_{\lambda_{2} \lambda'_{2}}
 (\vec{q}_{2},t) 
 \label{eq8}
\end{eqnarray}

where $u^T(\vec{q} \lambda | \vec{q}_{1}
\lambda_{1} ; \vec{q}_{2} \lambda_{2})$ and
$u^R(\vec{q} \lambda | \vec{q}_{1} \lambda_{1} ;
\vec{q}_{2} \lambda_{2})$ are  proportional to the
Clebsh-Gordon coefficients $C(l_{1} l_{2} l; 000)$ and
$C(l_{1} l_{2} l; 101)$, respectively. They vanish for
$l + l _{1} + l_{2}$ odd, and their explicit form is given
in \cite{10}. The important input into (\ref{eq8}) is the direct
correlation matrix $\mathbf{c} (\vec {q}) = (c_{\lambda \lambda '}
(\vec {q}))$ which is related to $\mathbf{S} (\vec
{q})$ by the Ornstein-Zernike equation for linear molecules.

\begin{equation} 
\label{eq9}
\mathbf{c}(\vec{q} ) = (4 \pi /\rho _0) [\mathbf{1-S}^{-1}
(\vec {q})] \qquad ,
\end{equation}

where $\rho _0 =N/V$ is the number density.

Since we want to investigate the dependence of $\mathbf{S}(\vec{q},t)$
on the aspect ratio it is important to check whether the
Eqs.~(\ref{eq4}) and Eq.(\ref{eq8}) reduce for $x_{0}=1$ to the
corresponding equations of motion for hard spheres. Due to the
complete isotropy of hard spheres it is:

\begin{equation}
\begin{array}{c} c_{\lambda \lambda'}(\vec{q})=
\end{array}
 \left\{
\begin{array}{cc}
4\pi c(\vec{q}), &(\lambda,\lambda')=(00,00) \\
\equiv 0, &(\lambda,\lambda')\neq(00,00)
\end{array}
\right.
\label{eq10}
\end{equation}

where $c(\vec{q})$ is the direct correlation function for hard
spheres. Substituting Eq.(\ref{eq10}) into Eq. (\ref{eq8}) strongly
reduces the number of mode-coupling terms. First of all one
obtains:

\begin{equation} \label {eq11}
m_{\lambda \lambda'}^{\alpha \alpha'} (\vec{q},t)\equiv 0 \, , 
\quad (\alpha, \alpha')\neq(T,T)
\end{equation}

Using $u^\alpha(\ldots)$ from ref.~\cite{10} we get for the
nonzero elements:

\vspace{1ex}
\begin{equation}
\label{eq12}
\begin{aligned} 
m_{\lambda \lambda'}^{TT}(\vec{q},t) & =  
\rho_{0}^{2} \, \frac {1}{2N}
\sum_{ \vec{q}_{1}, \vec{q}_{2} } \Big\{ \big( \frac{1}{q} 
\vec{q} \cdot \vec{q}_{1} \, c(\vec{q}_{1}) \big)^{2} 
S(\vec{q}_{1},t) 
S_{\lambda \lambda'} (\vec{q}_{2},t) \, + \\
& + \big( \frac{1}{q} \vec{q} 
\cdot \vec{q}_{2} \, c(\vec{q}_{2}) \big)^{2} S_{\lambda \lambda'}
(\vec{q}_{1},t) S(\vec{q}_{2},t) + 
\\[1ex]
& + \frac{1}{q^{2}}(\vec{q} \cdot \vec{q}_{1}) 
(\vec{q} \cdot \vec{q}_{2})
\, c(\vec{q}_{1}) \, c(\vec{q}_{2})
[ S_{00,\lambda'}(\vec{q}_{1},t) S_{\lambda,00}
(\vec{q}_{2},t) + (1 \leftrightarrow 2) ] \Big\}
\end{aligned}
\end{equation}

where we used $S(\vec{q},t)\equiv S_{00,00}(\vec{q},t)$, the
density correlator for hard spheres. Note that
$m_{00,00}^{TT}(\vec{q},t)$ coincides with the memory kernel for
simple liquids in general and for hard spheres in particular. But
$m_{\lambda \lambda'}^{TT} (\vec{q},t)$ seems to be nondiagonal in
$\lambda$. Such a nondiagonality would generate an
\textit{unphysical} influence (at least for hard spheres with
smooth surfaces) of $S_{\lambda \lambda} (\vec{q},t)$ with $\lambda
\neq (0,0)$ onto $S(\vec{q},t)$. However, using the initial
conditions Eq. (\ref{eq5}) and

\begin{equation} 
\label{eq13}
S_{\lambda\lambda'}(\vec{q})=0\quad,\quad\lambda\neq\lambda'
\end{equation}

which immediately follows from Eq. (\ref{eq9}) and Eq. (\ref{eq10})
one can prove that Eq. (\ref{eq4}) and Eq. (\ref{eq8}) yield:

\begin{equation} \label{eq14}
\frac{d^\nu}{dt^\nu}S_{\lambda\lambda'}(\vec{q},t)|_{t=0}=0
\quad,\quad\lambda\neq\lambda'
\end{equation}

for \textit{all} $\nu$. Although Eqs. (\ref{eq4}) and  (\ref{eq8}) 
are nonlinear we therefore expect
that Eq. (\ref{eq14}) implies

\begin{equation} \label{eq15}
S_{\lambda\lambda'} (\vec{q},t) \equiv
0\quad,\quad\lambda\neq\lambda'\quad,
\end{equation}

as for linear equations. In that case
$m_{\lambda\lambda'}^{\alpha\alpha'}(\vec{q},t)$ given by
Eq. (\ref{eq12}) becomes diagonal in $\lambda$, i.e. the
MCT-equations for the remaining correlators $S_{\lambda\lambda}
(\vec{q},t)$, $l\geq0$, decouple with respect
to $\lambda$. Therefore $S(\vec{q},t)$ is not influenced by
$S_{\lambda\lambda}(\vec{q},t)$ with $\lambda\neq(0,0)$ and it
obeys the MCT-equation for hard spheres. Because
$m_{\lambda\lambda}(\vec{q},t)$ for $\lambda\neq(0,0)$ contains
bilinear terms $S(\vec{q}_{1},t)S_{\lambda\lambda}(\vec{q}_{2},t)$ the
hard sphere correlator $S(\vec{q},t)$ has an influence on the
orientational correlators $S_{\lambda\lambda}(\vec{q},t)$, which
is unphysical, as well. The reason for this lies in our assumption
that the density modes $\rho_\lambda(\vec{q})$ are slow variables
for \textit{all} $\lambda$, which only will be justified for
$x_{0}\neq 1$. 
Neglecting $\mathbf{m}(\vec{q},t)$, Eqs. (\ref{eq4}) reduce
to a linear equation for the (normalized) correlators
$\mathbf{\Phi}(\vec{q,t})=\mathbf{S}^{-1/2}(\vec{q})
\mathbf{S}(\vec{q},t)\mathbf{S}^{-1/2}(\vec{q})$ \cite{18}:

\begin{equation} \label{eq16}
\ddot{\bg{\Phi}} (\vec{q},t) +\bg{\Omega}^{2}(\vec{q})
\bg{\Phi}(\vec{q},t)=0
\end{equation}

with the hermitean frequency matrix squared:

\begin{align} \label{eq17}
(\bg{\Omega}^{2} & (\vec{q}))_{lm,l'm'}
= \nonumber \\[1ex] 
& =\sum_{l''m''}(\mathbf{S}^{-1/2}
(\vec{q}))_{lm,l''m''} 
\left[\frac{k_{B}T}{M} q^{2}+\frac{k_{B}T}{I}
l''(l''+1) \right] (\mathbf{S}^{-1/2}
(\vec{q}))_{l''m'',l'm'}.
\end{align}

These microscopic frequencies will play an essential role for our
study of the microscopic dynamics. Since the static correlators
$S_{lm,l'm'}(\vec{q})$ for hard ellipsoids are nondiagonal there
will be a coupling between the various modes already on the linear
level of Eq. (\ref{eq16}). This type of coupling has been absent for
dipolar hard spheres due to the use of Wertheim's solution for
$c_{\lambda\lambda'}(\vec{q})$ for which the direct correlation
functions are diagonal in $l$ and $l'$ \cite{18}. Therefore in
contrast to our results for dipolar hard spheres we may expect
additional features for hard ellipsiods. The microscopic frequencies
will become "renormalized" close to the glass transition and particularly
deep in the glass. In ref.~\cite{18} it has been shown that the
microscopic frequencies in the glass follow from:

\begin{equation} \label{eq18}
(\bg{\widehat{\Omega}}^2(\vec{q}))^{\alpha\alpha'}=(\mathbf{J}^\alpha(\vec{q})^{1/2}
[\mathbf{q}^{\alpha} \mathbf{S}^{-1}(\vec{q})\mathbf{q}^{\alpha'}+\mathbf{C}^{\alpha\alpha'}(\vec{q})]
(\mathbf{J}^{\alpha'}(\vec{q}))^{1/2}
\end{equation}
with
$C_{\lambda \lambda'}^{\alpha \alpha'}(\vec{q})= \lim_{t \rightarrow \infty}
m_{\lambda \lambda'}^{\alpha\alpha'}(\vec{q},t)$.

\section{Results}

To solve Eqs.~(\ref{eq4}) and Eqs.~(\ref{eq8}) numerically one has to
cut-off $q$ and $l$. Since the numerical effort is considerable
for hard ellipsoids we have chosen $l\leq l_{co}=2$ and $30$
$q$-values which were distributed non-equidistantly between
$q_{min}\cong 0.51$ and $q_{co}=40$.

More details about the numerical procedure can be found in
ref.~\cite{18}. Furthermore we have restricted ourselves to
$x_{0}>1$. Although no strict symmetry exists between $x_{0}>1$ and
$x_{0}<1$ one may expect similar behavior for $x_{0}<1$. It would be
interesting to check this point. The direct correlation functions
needed as an input for the memory kernels were calculated from
Percus-Yevick theory \cite{21}. The temperature enters into
$\mathbf{J}^\alpha(\vec{q})$, only (cf. Eq.~(\ref{eq7})). We choose
$k_{B}T$ such that the thermal velocity $(k_{B}T/M)^{1/2}$ equals
one. This choice also sets the scale for the microscopic
frequencies $\Omega_{lm,l'm'}^{\alpha \alpha} \vec{q})$ and
$\widehat{\Omega}_{lm,l'm'}(\vec{q})$. All results will be given
for the $q$-frame.

\begin{figure}[http] 
\centerline{\rotatebox{0}{\resizebox{8cm}{!}{\includegraphics{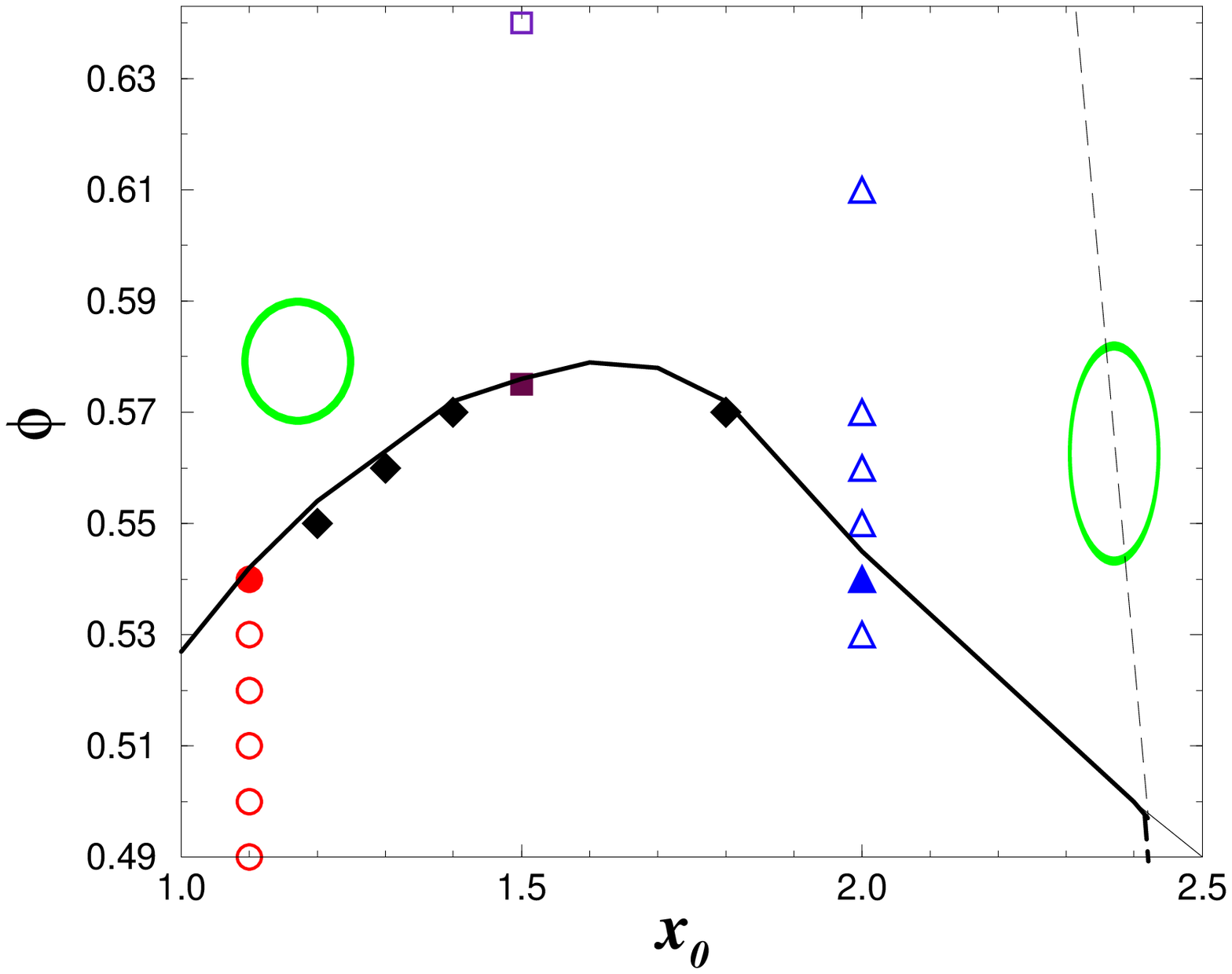}}}}
\caption{Glass transition phase diagram for hard ellipsoids: There are
two glass transition lines (solid and dashed lines). See text for
an explanation. The symbols show those points for which we have
calculated the time-dependent correlators. To demonstrate the
deviation from a hard sphere the cross section of an ellipsoid is
shown for $x_{0} =1.1$ and $x_{0} =2.4.$}\label{fig1}
\end{figure}

Before we come to the dynamical results we
present in figure \ref{fig1} the phase diagram. The liquid phase
is below the solid line and bounded by the thick part of the
dashed line. The remaining part corresponds to the glassy phase.
This diagram contains two glass transition lines which cross at $(\phi,
x_{0})\cong (0.498, 2.42)$. This crossing relates to a change of the
mechanism for the glass transition. For $1\leq x_{0}<2.42$ it is the
cage effect which mainly leads to a freezing into a glass whereas
for $x_{0}>2.42$ a precursor of a nematic order becomes responsible
for the glass transition \cite{12}. Note, that this change at $x_{0}
\cong 2.42$ does not happen suddenly but it is a crossover
phenomenon. Therefore the numerical value $x_{0} \cong2.37$ should
not be over-interpreted.

\begin{figure}[http] 
\centerline{\rotatebox{0}{\resizebox{10cm}{!}{\includegraphics{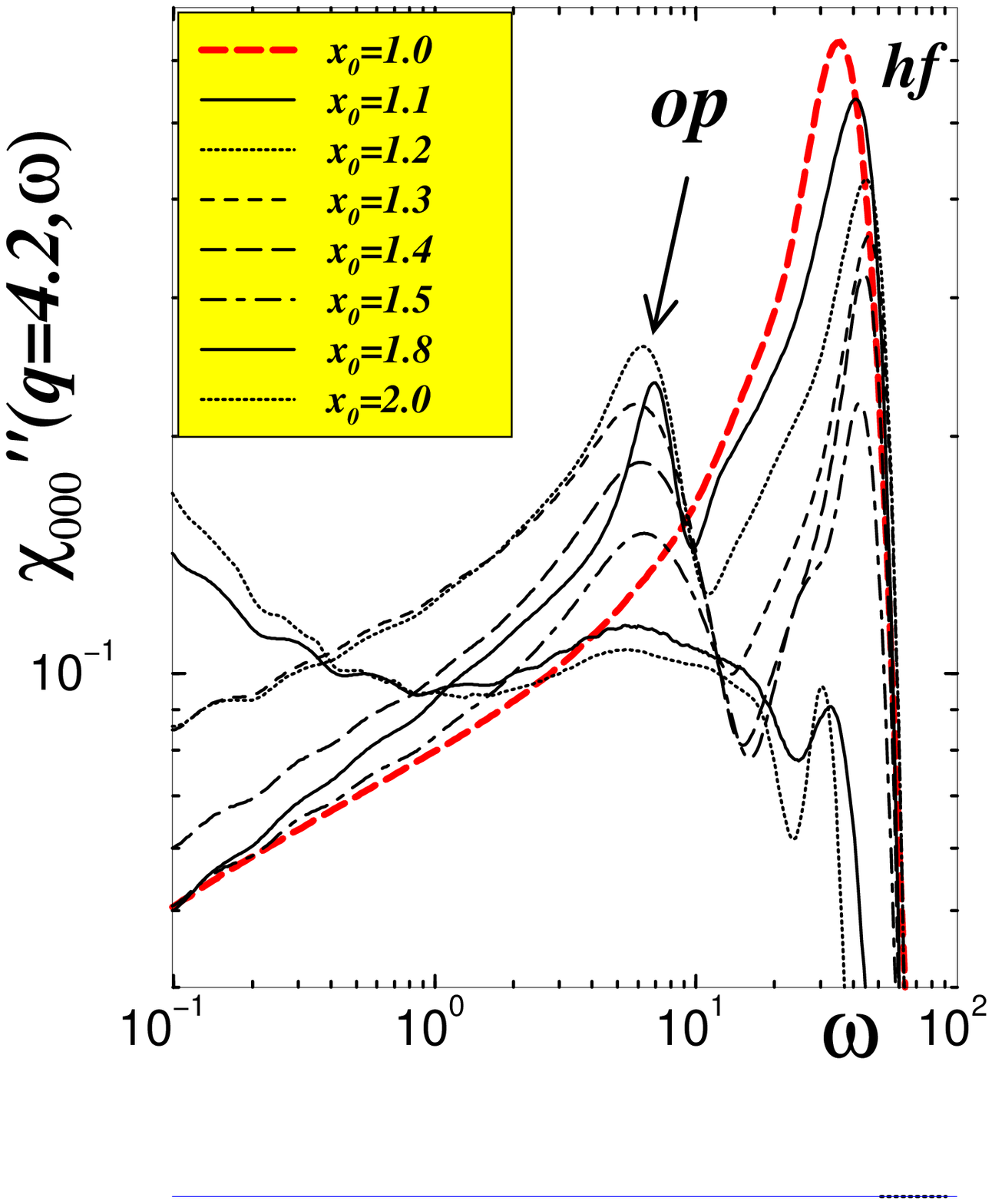}}}}
\caption{Compressibility $\chi^{''}_{000}(4.2,
\omega)$ for $(\phi,x_{0})$ indicated by the full symbols in figure
\ref{fig1}. The result for hard spheres $(x_{0}=1)$ with
$\phi=0.5184$ is shown by the thick dashed line.} 
\label{fig2}
\end{figure}

\begin{figure}[http] 
\centerline{\rotatebox{0}{\resizebox{10cm}{!}{\includegraphics{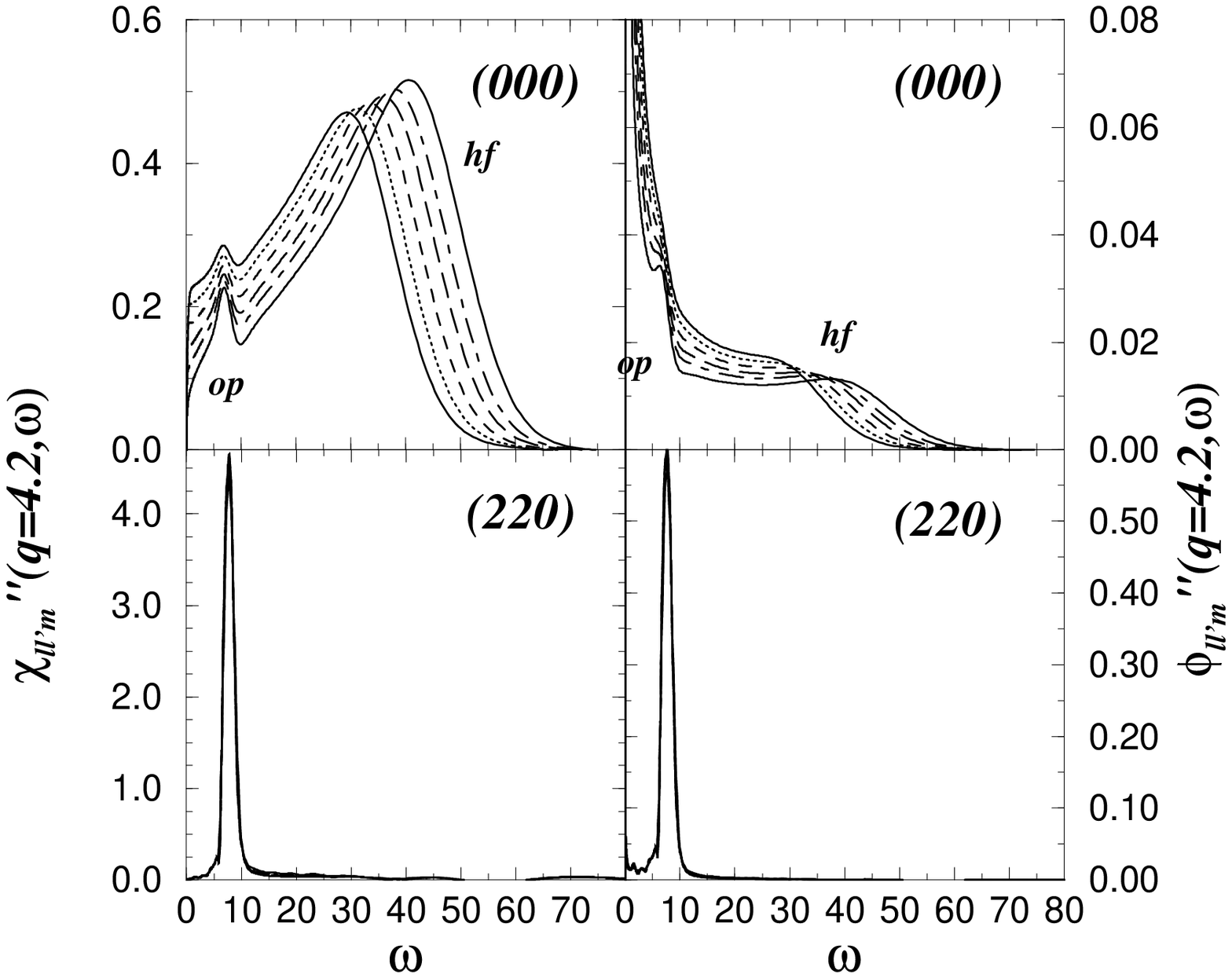}}}}
\caption{$\omega$-dependence of $\chi^{''}_{ll0}(4.2,
\omega)$ (left) and $\phi_{ll0}^{''}(4.2, \omega)$ (right) for
$l =0,2$, $x_{0}=1.1$ and the $\phi$-values marked in figure 1:
$\phi =$ \,0.49 (solid)\, 0.50 (dotted)\, 0.51 (short dashed) \, 0.52
(long dashed) \, 0.53 (chain) \, 0.54 (solid)} 
\label{fig3}
\end{figure}

The correlators $S_{ll'm}(q,t)$ for $l\leq l_{co}=2$ and
$q\leq q_{co} \cong 40$ were calculated from Eqs.~(\ref{eq4})
and ~(\ref{eq8}) for the points shown in figure \ref{fig1}. From
these results we have determined the correlation spectra
$\phi_{ll'm}^{''} (q,\omega)$ and the susceptibility
$\chi_{ll'm}^{''}(q,\omega)=\omega\phi_{ll'm}^{''}(q,\omega)$.
Figure \ref{fig2}  presents the compressibility spectrum for $q
\cong 4.2$ and pairs of $(\phi,x_{0})$ on the liquid side, but close
to the glass transition line (full symbols in figure \ref{fig1}).
Here we include $x_{0}=1.8$ for which the static correlators were
obtained from a MD-simulation \cite{13}. We have chosen
double-logarithmic scales on which the characteristic features
become more prominent. For hard spheres, i.e. $x_{0}=1$, only one
peak, which we called the high frequency peak (hf-peak), exists at
$\omega_{hf}$ for $\omega\geq 10^{-1}$. Now, making $x_{0}$ larger
than one, we observe an additional peak at $\omega_{op}$ about one
decade below the hf-peak. This peak is what we called the
orientational peak $(op)$ in the first section. It already appears
for $x_{0}=1.1$, the smallest value we have chosen besides $x_{0}=1$.
With increasing aspect ratio, up to our maximum value $x_{0}=2$,
the position of the $op$ almost does not change whereas
${\omega}_{hf}$ first increases and then decreases. The most
striking features are (i) the $op$ becomes broader and (ii) the
ratio of the height of the $op$ and the hf-peak increases with
increasing $x_{0}$. This behavior can be understood as follows.
Increasing $x_{0}$ enhances the steric hindrance which leads to a
stronger translational-rotational-coupling. Therefore the orientational
motion which is an almost free rotation with a single frequency
for $x_{0}$ close to one couples stronger and stronger to the
translational motion which enters into the compressibility.
This leads to a more hindered rotational motion involving a broader
band of rotational frequencies such that the $op$ becomes broader
and it leads to an enhancement of the orientational contribution
at $\omega_{op}$ relative to the translational one at $\omega_{hf}$.
Since $\omega_{hf}$ does not change with $I$ \cite{13} it must be
related to the translational motion. The fact that  a distinct
"fingerprint" of the orientational dynamics can be observed in the
compressibility spectrum already for $x_{0}=1.1$ might be surprising. But
figure \ref{fig3} supplies an explanation. There we show
$\chi_{ll'm}^{''}$ (4.2, $\omega$) and $\phi_{ll'm}^{''}$ (4.2, $\omega$)
on linear scales, for the volume fractions marked in
figure \ref{fig1} for $x_{0}=1.1$. Because these quantities for
$(ll'm)=(200), (221)$ and $(222)$ strongly resemble that for
$(ll'm)=(220)$ we only show data for $(000)$ and $(220)$.
$\chi_{220}^{''}$ $(4.2, \omega)$ and $\phi_{220}^{''}$ $(4.2,
\omega)$ exhibit one rather sharp peak at $\omega_{op}^{'} \approx
8$. For $x_{0}=1$ it follows from Eq.~(\ref{eq9}) 
with Eq.~(\ref{eq10}) that $S_{lm,l'm'}(\vec{q})=\delta_{ll'}
\delta_{mm'}$ for $l>0$. Using this, $q=0$ and $I=1/10$
(for $x_{0}=1$) we get from Eq.~(\ref{eq17}) the frequency
$\Omega_{2}^{\mathrm{free}}=(10 \cdot 2 \cdot 3)^{1/2} \cong 7.75$
for a free rotation of a sphere  with $l=l'=2$. Therefore it is obvious
that the peak in $\chi_{220}^{''}$ and $\phi_{220}^{''}$ at
$\omega_{op}^{'} \approx 8$ comes from an almost free rotation.
If the rotation is \textit{completely} free then
$\chi_{220}^{''}(q,\omega)$ and $\phi_{220}^{''}(q,\omega)$ are
proportional to $\delta (\omega - \Omega_{2}^{\mathrm{free}})$.
Although the coupling of the rotational motion into
$\chi_{000}^{''}$ is very small for $|1-x_{0}|\ll1$ it is the almost
$\delta$-like behavior of $\chi_{220}^{''}(q,\omega)$ and
$\phi_{220}^{''}(q,\omega)$ which can contribute significantly to
$\chi_{000}^{''}(q,\omega)$ and $\phi_{000}^{''}(q,\omega)$. This
contribution is more apparent for $\chi_{000}^{''}$ than for
$\phi_{000}^{''}$, as can be seen from figure \ref{fig3}. Whereas the
$op$ at $\omega_{op}\approx\Omega_{2}^\mathbf{free}$ occurs in
$\chi_{000}^{''}$ for all studied volume fractions it
only becomes present in $\phi_{000}^{''}$ closer to the glass
transition. The fact that $\omega_{op}\approx \omega_{op}^{'}$
supports the rotational origin of the $op$. Two more features follow. 
First, the spectra for $(ll'm)=(000)$ depend sensitively on    
$\phi$ whereas those for $(ll'm)=(220)$ almost do not
change, and second $\omega_{op}$ is practically constant in
contrast to the position $\omega_{hf}$ of the hf-peak which moves
towards higher frequencies with increasing $\phi$. The latter 
observation implies that with increasing $\phi$ the glass becomes
more stiff with respect to the translational than for the rotational
dynamics, provided $x_{0}$ is close to one.

\begin{figure}[http] 
\centerline{\rotatebox{0}{\resizebox{10cm}{!}{\includegraphics{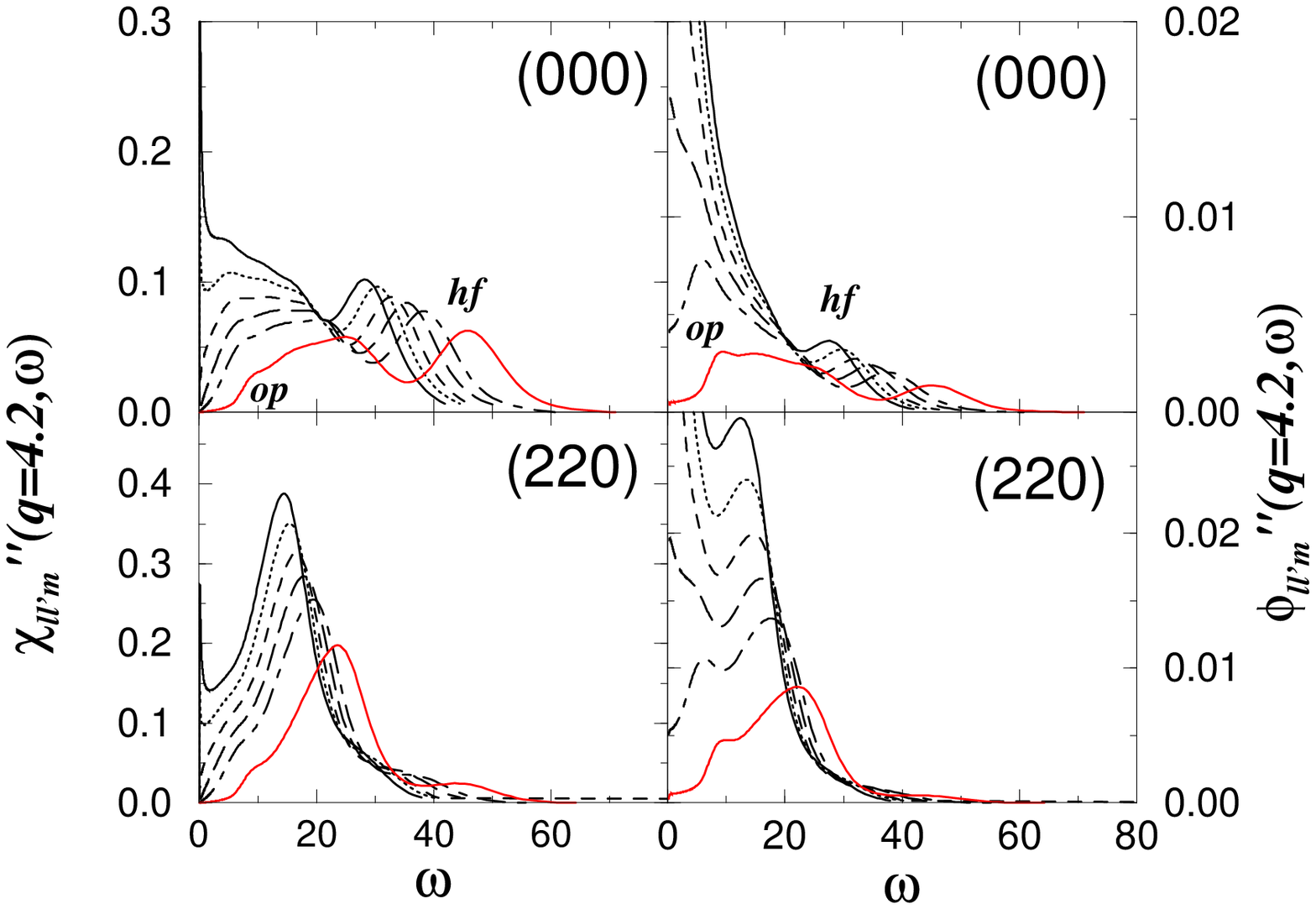}}}}
\caption{$\omega$-dependence of $\chi_{ll0}^{''}(4.2,
\omega)$ (left) and $\phi^{''}_{ll0}(4.2, \omega)$
(right) for $l= 0,2\,, x_{0}=2$ and the $\phi$-values marked in
figure \ref{fig1}: $\phi$ = 0.53 (solid) \, 0.54 (dotted) \,
0.55 (short dashed) \, 0.56 (long dashed) \, 0.57 (chain) \,
0.61 (solid)} 
\label{fig4}
\end{figure}

Now let us consider the results for the largest value for $x_{0}$ we
have studied. Figure \ref{fig4} presents the spectra (again on
linear scales) for $x_{0}=2$ and the volume fractions indicated in
figure \ref{fig1}. The glass translation occurs at
$\phi_c(x_{0}=2)\cong 0.545$. Comparison of figure \ref{fig4} with
figure \ref{fig3} shows that all the spectra have become more
structured and broader due to the increase of $x_{0}$. The rather
narrow peak in $\chi_{220}^{''}$ $(4.2, \omega)$ and $\phi_{220}^{''}$
$(4.2, \omega)$ at $\omega_{op}^{'}\approx 8$ for $x_{0}=1.1$ has almost
disappeared. It occurs as a shoulder in $\chi_{220}^{''}$  and as a
peak in
$\phi_{220}^{''}$
deeper in the glass. Its shape and intensity depends on
$\phi$. A well pronounced peak also occurs in $\chi_{220}^{''}$ and 
$\phi_{220}^{''}$ at $\omega_{int}$ between $\omega\approx 15$ and
$\omega\approx 25$, depending on $\phi$. This peak is present in
the liquid $(\phi\leq0.54)$ and the glass $(\phi\geq 0.55)$.

These two peaks have their counterparts in $\chi_{000}^{''}$
and $\phi_{000}^{''}$ at $\omega_{op}\approx\omega_{op}^{'}$ and
$\omega_{int} \approx \omega_{int}^{'}$, although
the peak in $\chi_{000}^{''}$ at $\omega \approx
\omega_{int}$ becomes more pronounced deeper in the glass,
and has a shoulder-like behavior for $\phi_{000}^{''}$, only. Besides
these two peaks a hf-peak exists above $\omega\approx 25$, for all
volume fractions. We have checked the dependence of these peak
positions on the moment of inertia $I$ and have found that only
$\omega_{op}\approx \omega_{op}^{'}$ shifts with $I$. Therefore the
peak at $\omega_{op}\approx \omega_{op}^{'}$, which is much less
pronounced then for $x_{0}=1.1$ is the orientational peak. The
origin of the intermediate peak will be elucidated below.

\begin{figure}[http] 
\centerline{\rotatebox{0}{\resizebox{10cm}{!}{\includegraphics{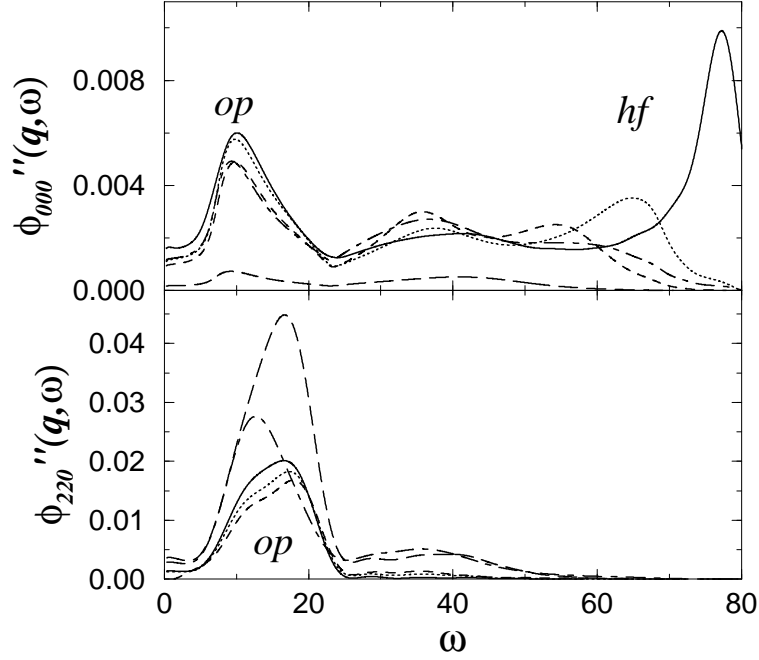}}}}
\caption{$\omega$-dependence of $\phi^{''}_{l
l0}(q,\omega)$ for $l=0,2 \; , x_{0}=1.5, \phi = 0.64$ and
$q=3.1$ (solid line), 4.2 (dotted line), 4.7 (short dashed line),
6.5 (long dashed line) and 9.8 (chain line)} 
\label{fig5}
\end{figure}

It is also interesting to study the $q$-dependence of the spectra.
This has been done for an intermediate aspect ratio $x_{0}=1.5$ and
$\phi=0.64$, i.e. deep in the glass (cf. figure \ref{fig1}). The
result is depicted in figure \ref{fig5} for
$\phi_{ll0}^{''}(q,\omega)$ and $l=0,2$. The $q$-dependence of
$\chi_{ll0}^{''}(q,\omega)$ is quite similar. Even more pronounced
than for $x_{0}=2$ we observe two peaks in $\phi_{220}^{''}$ at
$\omega_{op}^{'}\approx 10-20$ and at $\omega_{int}^{'}\approx
35$. Again, these two peaks have their counterpart in
$\phi_{000}^{''}(q,\omega)$ at $\omega_{op}$ and
$\omega_{int}$. In addition there is a hf-peak at $\omega_{hf}$.
The dispersion of these peak positions depends on $l$. $\omega_{op}$
is practically $q$-independent whereas $\omega_{op}^{'}$ exhibits
some $q$-dependence, and the position of the intermediate and the
hf-peak shows dispersion for $l = 0$ and $l= 2$.

\begin{figure}[http] 
\centerline{\rotatebox{0}{\resizebox{10cm}{!}{\includegraphics{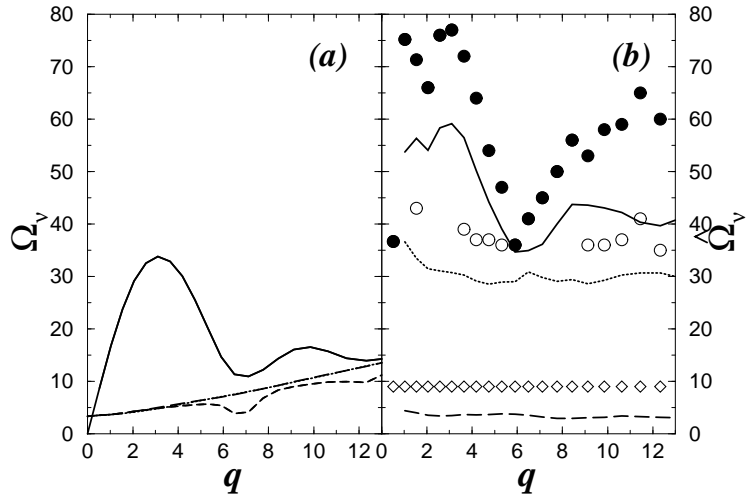}}}}
\caption{$q$-dependence of (a) $\Omega_{3}^{+}$ (solid
line), $\Omega_{3}^{-}$ (short dashed line), $\Omega_{1}$ (long dashed
line) and $\Omega_{2}$ (dotted line) (see text). (b)
$\hat{\Omega}_{1}$ (dashed line), $\hat{\Omega}_{2}$ (dotted line) and
$\hat{\Omega}_3$ (solid line) (see text). The peak positions
$\omega_{op}$ (diamonds), $\omega_{int}$ (open circles) and
$\omega_{hf}$ (full circles) were deduced from
$\phi^{''}_{000}(q,\omega)$ for $x_{0}=1.5$ and $\phi = 0.64$.}
\label{fig6}
\end{figure}

To understand the existence and the $q$-dependence of these peaks
at least qualitatively we have calculated for $x_{0} =1.5, \; \phi =
0.64$ and $l \leq l_{co} =2$ the eigenfrequencies
$\Omega_3^{\pm} (q)$ of the $2\times 2$ matrix
$(\Omega_{l0,l' 0}(q))$, $\Omega_{1}(q) \equiv
\Omega_{21,21}(q)$, $\Omega _{2}(q) \equiv \Omega_{22,22}(q)$ and
the "renormalized" frequencies $\hat{\Omega}_\nu (q)$, $\nu = 1,2$
and $3$ which are the eigenfrequencies of the $3 \times 3$ matrix
$(\hat{\Omega}^{\alpha \alpha '}_{l0,l'0} (q))$ (cf.
\cite{18}). Since $\hat{\Omega}^{RR}_{l0,l'0}(q)=0$ for
$l$ and/or $l'$ equal to zero, $(\hat{\Omega}^{\alpha \alpha
'}_{l0, l'0} (q))$ is indeed a $3 \times 3$ matrix for
$l,l' \leq 2$. The results for $\Omega _\nu (q)$ and
$\hat{\Omega} _ \nu (q)$ are given in figure \ref{fig6}a and
\ref{fig6}b, respectively. Figure \ref{fig6}b also contains
$\omega _{op}(q)$, $\omega _{int}(q)$ and $\omega _{hf} (q)$
determined from $\phi^{''}_{000} (q,\omega)$ for $x_{0}=1.5$ and
$\phi=0.64$, i.e. deep in the glass. Since $\Omega_{1}(q)$ cannot
be distinguished from $\Omega_{2}(q)$ there are practically three
branches in figure \ref{fig6}a. Because $\Omega_3^+(q)\rightarrow 0$
for $q \rightarrow 0$ it is the longitudinal acoustic branch, whereas the
others are optic-like. These optic-like behavior also appears for
the "renormalized" frequencies $\hat{\Omega} _{1}(q)$ and
$\hat{\Omega}_{2}(q)$ which seem to reproduce at least qualitatively
the results for $\omega_{op}(q)$ and $\omega_{int}(q)$. The same holds for
$\omega_ {hf}(q)$ which is in phase with $\hat{\Omega}_3(q)$ where
$\hat{\Omega}_3(q)$ is the "renormalized" bare frequency $\Omega
^+_3 (q)$. Unfortunately, our results for $\hat{\Omega}_{\nu}(q)$
become less accurate for small $q$. For an expalantion see ref. \cite{18}.
Therefore we cannot definitely determine the dispersion for $q\rightarrow 0$.
The intermediate peak only appears deep in the glass. In
that case the nondiagonal static correlator $S_{200}(q)$ becomes
large. Therefore we believe that the origin of that peak comes
from the additional coupling between the modes with $l= 0$ and
$l= 2$ which already exists on the linear level (see
discussion at the end of section 2).

\section{Summary and conclusions}

We have calculated the dynamics of a system of hard ellipsoids as
a function of the volume fraction $\phi$ and the aspect ratio
$x_{0}$. In the spectra for the center of mass dynamics we have
found an orientational peak at $\omega_{op}$, roughly one decade
below a high-frequency peak. Due to its shift with the moment of
inertia and its appearance in the spectra of the "quadrupolar"
motions (quantities with $\ell =2$) its origin is mainly of
orientational nature. This peak becomes broader and its intensity
relative to the intensity of the hf-peak increases with increasing
$x_0$, i.e. with the increase of the steric hindrance. Its
position is practically q-independent and it scales with
$I^{-1/2}$. This implies that it originates mainly from localized
modes with $q \approx 0$, as already found in refs. \cite{13} and
\cite{18}. For the dipolar hard sphere system studied in ref.
\cite{18} it has also been speculated that this orientational peak
could be one of several contributions to the \textit{boson
peak}. Since hard ellipsoids are an athermal system we can not
draw such a conclusion here. However, the appearance of the $op$ for
all aspect ratios we have studied together with its existence for
dipolar hard spheres may give a hint that it may occur in many
molecular liquids. That it is not an artefact of the approximate
character of the used equations of motions has clearly been
demonstrated by the MD-simulation \cite{13}.

It is also interesting for $x_{0}$ close to one that a variation of
$\phi$ almost does not affect the spectra for $l= l' = 2$,
in contrast to those for $l= l' = 0$. There is a
significant influence of the rotational motion on $\chi_{000}^{''}
$ and $\phi_{000}^{''} $ but much less feedback from the
translational motion to $\chi_{220}^{''} $ and $\phi_{220}^{''} $
for weak steric hindrance.

Finally, deep in the glass we have found an intermediate peak
between the $op$ and hf-peak. This peak did not appear for dipolar
hard spheres where the static correlators were taken diagonal in
$l$ and $l'$. Since this is not the case for the system of
hard ellipsoids we believe that the intermediate peak turns up
due to a coupling between modes with $l= 0$ and $l=2$
mediated by the nondiagonal elements of $S_{ll ' m}(q)$.
The q-dependence of the orientational, intermediate and the
hf-peak in the compressibility spectrum can be described by the
"renormalized" eigenfrequencies, at least on a qualitative level.

\bigskip

\textbf{Acknowledgement}: We gratefully acknowledge financial support by
the "Sonderforschungsbereich 262" (Deutsche
Forschungsgemeinschaft) of the present research project and of
those during the last ten years.

\eject

\newpage


\end{document}